\documentclass[final,3p,times,twocolumn]{elsarticle}
 \biboptions{comma,sort&compress}
\usepackage{here}
 \usepackage{graphicx}
  \usepackage{epsfig}

\def\nin{\noindent}
\def\beq{\begin{equation}}
\def\eeq{\end{equation}}
\def\bea{\begin{eqnarray}}
\def\eea{\end{eqnarray}}

\journal{Nuc. Phys. (Proc. Suppl.)}

\begin{document}

\begin{frontmatter}



\title{Symmetry breaking effect on determination of polarized and unpolarized parton distributions}

\author[label1,label2]{F. Arbabifar}
\address[label1]{Physics Department, Semnan University, Semnan, Iran}
\ead{farbabifar@ipm.com}

\author[label1,label2]{Ali. N. Khorramian \corref{cor1}}
\address[label2]{School of Particles and Accelerators, IPM (Institute for Studies \\ in
        Theoretical Physics and Mathematics), P.O.Box 19395-5531,
        Tehran, Iran\\}\cortext[cor1]{Speaker}
\ead{Khorramiana@theory.ipm.ac.i}

\author[label1,label2]{H. Khanpour }

\ead{Hamzeh.Khanpour@nit.ac.ir}

\author[label2]{S. Atashbar Tehrani}

\ead{Atashbar@ipm.ir}

\begin{abstract}
\noindent We perform a new extraction for unpolarized and polarized
parton distribution functions considering a flavor decompositions
for sea quarks and applying very recent deep inelastic scattering
(DIS) and semi inclusive deep inelastic scattering (SIDIS) data in
the fixed flavor number scheme (FFNS) framework. In the new symmetry
breaking scenario the light quark and antiquark densities are
extracted separately and new parametrization forms are determined
for them. The heavy flavors contribution, including charm and bottom
quarks, are also taken to be account for unpolarized distributions.

\end{abstract}


\end{frontmatter}


\section{Introduction}
\nin
In the recent years our knowledge about the  structure of the  nucleons
 has improved and by the increase of both acceptable accuracy and the volume of data
 from deep inelastic scattering processes, new investigations
are also in remarkable progress \cite{DeRoeck:2011na,Boer:2011fh}.\\
In DIS experiments the photon transfers the electron vertex momentum
to the proton and scatters off spin-$\frac{1}{2}$, pointlike quark
component of it. The probability that the parton of flavor $f$
carries fraction $x$ of the struck proton momentum is called parton
distribution functions (PDFs) and plays a very important role to
determine DIS cross sections. The extraction of PDFs and polarized
PDFs (PPDFs) is developed to very precise QCD  analysis in
next-to-leading order (NLO) or even next-to-next-to-leading order
(NNLO) approximation which are based on new model
independent hypotheses \cite{Tung:2006tb,Lai:2010vv,Martin:2009bu,Martin:2010db,alt98abfr,Ball:1997sp,bou98,Atashbar Tehrani:2007be,Khorramian:2006wg,Khanpour:2011zz,Khanpour:2012zz,Arbabifar:2011zz,Khorramian:2010zz}.\\
\noindent The inability of inclusive DIS data to distinguish quarks
from antiquarks was always the main reason of symmetry consideration
by many theoretical groups until very recent years and now the
growing of SIDIS for polarized and Drell-Yan experiments for
unpolarized
data \cite{Towell:2001nh,Webb:2003ps,hermespdf,compasspdf} is the motivation of considering symmetry breaking models in the new analysis. \\
 In our previous works we studied the impact of the recent very
precise inclusive structure functions data from DIS experiments on
the determination of unpolarized and polarized parton distributions
in the standard scenario i.e. $\bar{u}=\bar{d}=\bar{s}$
\cite{PRD11,Khanpour:2010zz}, and now we apply SU(2) and SU(3)
symmetry breaking scenario and have $\bar{u}\neq\bar{d}\neq\bar{s}$
in determination of PDFs and PPDFs like what other groups have
studied recently
\cite{Broadhurst:2004jx,lss,Kataev:2004wv,deFlorian:2009vb}. In the
current analysis we apply
 experimental data of  inclusive Neutral Current Deep Inelastic Scattering
(NC DIS) for unpolarized and
spin-dependant DIS and SIDIS data for polarized QCD fit process explained in detail in \cite{Khanpour2012,DSPIN11}.\\
 The organization of the present paper is as follows: determination of
unpolarized PDFs is presented in Sec.~2 and polarized PDFs
extraction is discussed in Sec.~3. Finally in Sec.~4 we summarize
and give the conclusion of the analysis.
\section{Determination of unpolarized parton distributions}
\begin{figure}[h]
\includegraphics[width=80mm]{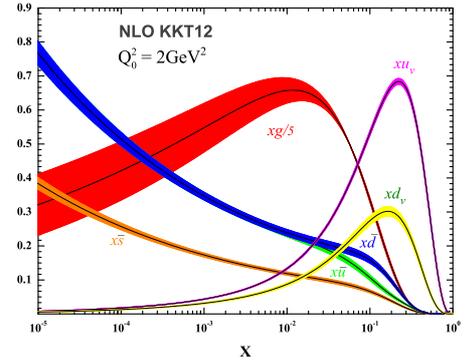}
 \caption{The {\tt KKT12} parton distribution functions as a function
of $x$ at initial scale $Q_0^2$ = 2 GeV$^2$ in NLO
approximation.}\label{nlopdf}
\end{figure}

\begin{figure}[h]
\vspace{0.5cm}
\includegraphics[width=80mm]{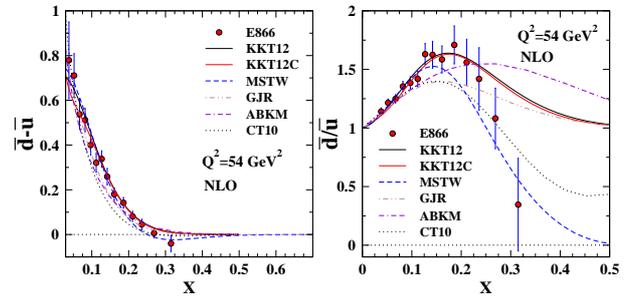}
\caption{Our results for $\bar{d}-\bar{u}$ and $\bar{d}/\bar{u}$ as
a function of $x$ in comparison to the results from CT10
\cite{Lai:2010vv}, MSTW08 \cite{Martin:2009iq}, ABKM10
\cite{Alekhin:2009ni} and GJR08 \cite{Gluck:2007ck}. The E866
results \cite{Towell:2001nh,Webb:2003ps}, scaled to Q$^2$=54
GeV$^2$, are also shown as circles.} \label{bar}
\end{figure}

\nin
The total structure function of proton $F_2^p (x,Q^2)$ in $\overline{\rm MS}$ factorization scheme can be written in NLO approximation as
\cite{Martin:2009iq}
\begin{eqnarray} \label{eq1}
F_2(x,Q^2) &=& F_{2,{\rm NS}}^+(x,Q^2) + F_{2,S}(x,Q^2) \nonumber \\
 &+& F_2^{(c,b)}(x,Q^2,m_{c,b}^2)\,,
\end{eqnarray}
here the non--singlet contribution is given by
\begin{eqnarray}
\frac{1}{x}\, F_{2,{\rm NS}}^+(x,Q^2) & = &
\Big[C_{2,q}^{(0)} + \frac{\alpha_{s}}{4\pi} C_{2,{\rm NS}}^{(1)} \Big]  \nonumber \\
& \otimes& \left[\frac{1}{18}\, q_8^+ +\frac{1}{6}\, q_3^+\right](x,Q^2)\,,
\end{eqnarray}
and the flavor singlet contribution is
\begin{eqnarray}
\frac{1}{x}\, F_{2,S}(x,Q^2)&=&\frac{2}{9} \left\{ \left[ C_{2,q}^{(0)} + \frac{\alpha_{s}}{4\pi} C_{2,q}^{(1)} \right]
\otimes \Sigma \right. \nonumber \\
&+& \left.  \, \frac{\alpha_{s}}{4\pi} C_{2,g}^{(1)} \otimes g\right\} (x,Q^2) .
\end{eqnarray}
The contribution of heavy flavors $F_{2} ^{c,b} (x,Q^2)$  have been
also added in our analysis and they are taken as in
Ref.~\cite{Gluck:2006pm}. In the above equations $\alpha_s$ is the
strong coupling constant, $C_{2,q}^{(0)}(z)=\delta(1-z)$,
$C_{2,q}^{(1)}=C_{2,\rm NS}^{(1)}$ and the additional NLO
$C_{2,g}^{(1)}$  and $C_{2,{\rm NS}}^{(1)}$ are the corresponding
known Wilson coefficients which can be found in
Ref.~\cite{vanNeerven:2000uj}. The PDFs combinations of $q_3^+$ and
$q_8^+$ and $\Sigma(x,Q^2)$  are also well determined in the literatures \cite{Khanpour2012}.\\
Here we consider symmetry breaking for
$\bar{u}\neq\bar{d}\neq\bar{s}$ and  a symmetry for strange sea,
$s=\bar{s}$, so our analysis is effected by these new assumptions.
For our QCD fit  we use the following parametrization forms of the
parton distribution functions at the input initial scale $Q_0^2$=2
GeV$^2$
\begin{eqnarray}
\hspace{-5mm}
xu_v   &=& A_{u_v}\, x^{\alpha_{u_v}}(1-x)^{\beta_{u_v}}(1+\gamma_{u_v}\, x^{\delta_{u_v}} + \eta_{u_v}\, x),  \nonumber \\
\hspace{-5mm}
xd_v    &=& A_{d_v}\, x^{\alpha_{d_v}}(1-x)^{\beta_{d_v}}(1+\gamma_{d_v}\, x^{\delta_{d_v}} + \eta_{d_v}\, x),  \nonumber \\
\hspace{-5mm}
x\Delta &=& A_{\Delta}\, x^{\alpha_{\Delta}}(1-x)^{\beta_S + \beta_{\Delta}}
(1+\gamma_{\Delta}\, x^{\delta_{\Delta}} + \eta_{\Delta}\, x),  \nonumber \\
\hspace{-5mm}
xS     &=& A_S\, x^{\alpha_S}(1-x)^{\beta_S}(1+\gamma_S\, x^{\delta_S} + \eta_S\, x),    \nonumber \\
\hspace{-5mm}
xg     &=& A_g\, x^{\alpha_g}(1-x)^{\beta_g}(1 +\gamma_g\, x^{\delta_g} + \eta_g\, x ), \label{PDFs}
\end{eqnarray}

\noindent here we take $x\Delta = x (\bar{d} - \bar{u} )$, $xS =
2x(\bar{u}+\bar{d}+\bar{s})$ and as we mentioned above $s= \bar{s}$,
since our used data sets are not sensitive to the special choice of
the strange sea parton distributions. Due to applying some
reasonable constraints in the parameter space of our global QCD fit
\cite{Khanpour2012}, only 13 parameters remained free for all parton
flavor in the final minimization. The $\chi_{\mathrm{\rm
global}}^{2}$ in global fit procedure minimization is defined
as~\cite{Martin:2009iq}

\begin{eqnarray} \label{chi2}
\hspace{-8mm}
\chi_{\mathrm{\rm global}}^{2}= \sum_{i=1}^{n^{\rm data}} \left[\left(\frac{{\cal N}_{i} - 1}{\Delta{\cal N}_{i}}\right)^{2}
+\sum_{j=1}^{n^{\rm data}} \left( \frac{{\cal N}_{j} D_{j}^{\rm data} - T^{\rm theory}_{j}} {{\cal N}_{j} ~ \Delta D_{j}^{\rm data}} \right)^{2}\right],
\end{eqnarray}
where $n^{data}$ shows the number of included data points and
$D_{i}^{\rm data}$, $\Delta D_{i}^{\rm data}$, and $T_{,i}^{\rm
theory}$ are the value, uncertainty and theoretical value for the
$n^{\mathrm{th}}$ data point of the $i^{\mathrm{th}}$ experiment.
${\Delta{\cal N}_{n}}$ is known as the experimental normalization
uncertainty and the value of ${\cal N}_{n}$ shows an overall
normalization factor for the $n^{th}$ experiment data. In our global
fits, we get $\frac {\chi^{2}}{NDF} =1.098$
and for the total number of used data points we put $n^{data}=3279$ introduced in Ref.~\cite{Khanpour2012}.\\
Our analysis process is accomplished using the QCD-PEGASUS package
in the fixed-flavor number scheme with consideration of massless
partonic flavors and $N_{f}=3$ \cite{Vogt:2004ns}. The results of
fitted parton distribution functions, known as {\tt KKT12}, and
their errors at the initial scale are presented  in
Fig.~\ref{nlopdf} and regarding to the symmetry breaking scenario, a
comparison of our results for $\bar{d}-\bar{u}$ and
$\bar{d}/\bar{u}$ as a function of $x$, with the results from other
groups and experimental data is shown in Fig.~\ref{bar} in NLO
approximation.

%
\section{Determination of polarized parton distributions}
\begin{figure}
\includegraphics[width=80mm]{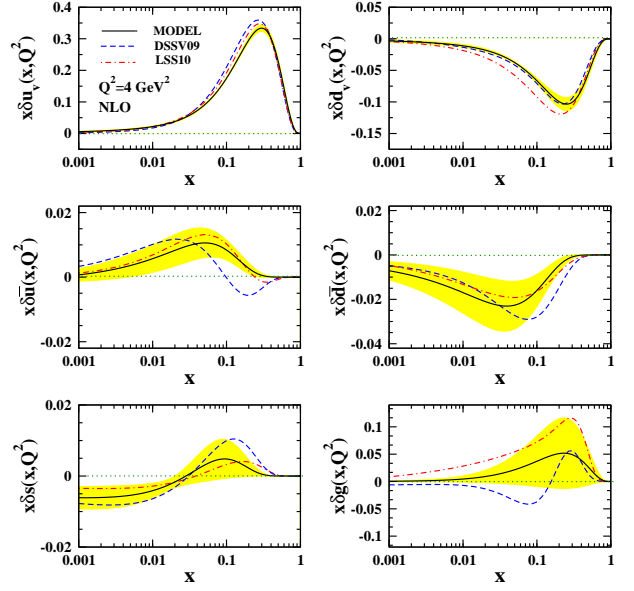}
 \caption{The quark polarized distributions as function of $x$ comparing with DSSV09
\cite{deFlorian:2009vb} and LSS10 \cite{lss} models  in NLO approximation.}\label{pdf}
\end{figure}

\begin{figure}[ht]
\hspace{5mm}
\includegraphics[width=58mm]{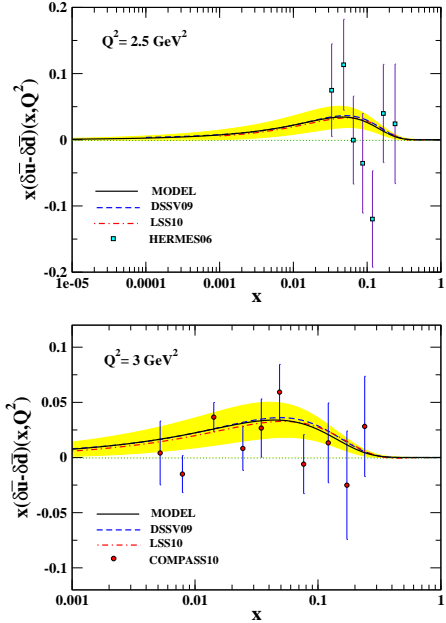}
 \caption{The quark polarized
distributions for the difference of $x(\delta\bar{u}-\delta\bar{d})$
at $Q^2=2.5~,3$ GeV$^2$ comparing to results from other models and
experimental data
\cite{deFlorian:2009vb,lss,hermespdf}.}\label{mines}
\end{figure}

\nin Generally we consider a nucleon is formed of massless partons
that have negative and positive helicity distributions $q_{\pm}(x,
Q^2)$ and the difference
\begin{equation}
\delta q(x, Q^2) = q_{+}(x, Q^2) - q_{-}(x, Q^2)\ ,
\end{equation}
shows how much the parton $q$ is responsible for the original
proton polarization and is called polarized parton distribution function.\\
In the present analysis for PPDFs determination  we subjoin  very
recent SIDIS experimental data for polarized parton densities from
HERMES \cite{hermespdf} and COMPASS \cite{compasspdf} to DIS
experimental data of Ref.~\cite{PRD11} since these additional
experiments help us to apply
symmetry breaking and recognize $\bar{u}$, $\bar{d}$ and $\bar{s}$ separately.\\
The polarized structure function $g_1(x,Q^2)$ is written in terms of
a Mellin convolution of  PPDFs with the relevant known Wilson
coefficients $\Delta C_{q,g}$ \cite{Lampe:1998eu}
\begin{eqnarray}
\hspace{-0.75cm}
\label{eqg1} g_1(x,Q^2)&=& \frac{1}{2} \sum_{q=u,d,s} e_q^2
\left\{ \left[1+\frac{\alpha_{s}}{2\pi}\Delta C_{q}\right]\right.
 \otimes \left[\delta q+\delta\bar{q}\right]\nonumber \\
 & + & \left.\frac{\alpha_{s}}{2\pi}\:2\Delta C_{g}\otimes\delta g\right\}\left({x },Q^2\right) ,
\end{eqnarray}
where $\alpha_s$ is the strong coupling constant, $e_q$ shows the charge of the quark flavor $q$ and $\{\delta
q,\delta\bar{q},\delta g\}$ are the corresponding PPDFs.
For our analysis we choose following functional forms for polarized PDFs in the initial scale
$Q_{0}^2$ = 4 GeV$^2$
\begin{eqnarray}
\hspace{-5mm}
x\:\delta u_v&=&{\cal
N}_{u_v}\eta_{u_v}x^{a_{u_v}}(1-x)^{b_{u_v}}(1+d_{u_v}x),\nonumber\\
\hspace{-5mm}
x\:\delta d_v&=&{\cal
N}_{d_v}\eta_{d_v}x^{a_{d_v}}(1-x)^{b_{d_v}}(1+d_{d_v}x),\nonumber\\
\hspace{-5mm}
x\:\delta \Delta& =&{\cal
N}_\Delta\eta_\Delta x^{a_\Delta}(1-x)^{b_\Delta} (1+c_\Delta\sqrt{x}),\nonumber\\
\hspace{-5mm}
x\:\delta \Sigma &=&{\cal N}_\Sigma \eta_\Sigma x^{a_\Sigma}(1-x)^{b_\Sigma}(1+c_\Sigma\sqrt{x}),\nonumber\\
\hspace{-5mm}
x\:\delta s &=&{\cal
N}_{s}\eta_{s}x^{a_{s}}(1-x)^{b_{s}}(1+d_{s}x),\nonumber\\
\hspace{-5mm}
 x\:\delta
g&=&{\cal N}_{g}\eta_{g}x^{a_{g}}(1-x)^{b_{g}},
\label{eq:parm}\end{eqnarray}
 \noindent where $\delta \Delta=\delta\bar{d}-\delta\bar{u}$ and $\delta \Sigma=\delta\bar{d}+\delta\bar{u}$. The normalization
constants ${\cal N}_{q}$ are determined such that the value of
$\eta_{q}$ become the first moments of PPDFs, i.e.
$\eta_{q}=\int_{0}^{1}dx\delta q(x,Q_{0}^{2})$. Since the current
SIDIS data are not sufficient yet to differ $s$ from $\bar{s}$ , we
apply $\delta{s}=\delta{\bar{s}}$ throughout, also we have to
 make some constraints on the parameter space to control the $x$ dependance of PPDFs \cite{DSPIN11}  like what we do for unpolarized PDFs.\\
The value of parameters $\eta_{u_{v}}$ and $\eta_{d_{v}}$ shows the
first moments of polarized valence quark distributions $\delta
u_{v}$ and $\delta d_{v}$ which can be linked to $F$ and $D$
determined in neutron and hyperon $\beta$--decays \cite{PDG} by
assuming $SU(2)$ and $SU(3)$ flavor symmetries
\cite{deFlorian:2009vb}. These quantities result into
$\eta_{u_{v}}=+0.928\pm0.014\ $ and $\eta_{d_{v}}= -0.342\pm0.018$
as shown in Ref.~\cite{PRD11}. \noindent Since in the present
analysis we are not interested to force $SU(2)$ and $SU(3)$ flavor
symmetry, we should relax the symmetry relations in $\eta_{u_v,d_v}$
measurements by introducing two flexible parameters,
$\varepsilon_{_{SU(2)}}$ and $\varepsilon_{_{SU(3)}}$ like what
DSSV09 \cite{deFlorian:2009vb} has proposed
\begin{eqnarray}
\hspace{-7mm}
\Delta \Sigma _{u}-\Delta\Sigma_{d}&=& \left( F+D\right) \left[ 1+\varepsilon_{_{SU(2)}}\right],\label{mom1}\\
\hspace{-7mm}
\Delta \Sigma_{u} +\Delta\Sigma_{d}-2\Delta\Sigma_{s}&=&
\left(3F-D\right)\left[ 1+\varepsilon_{_{SU(3)}} \right].\
\label{mom2}
\end{eqnarray}
 In above equations $\varepsilon_{_{SU(2,3)}}$ determine the deviation value from $SU(2)$ and $SU(3)$ symmetries and are also considered in the QCD global fit as free parameters.\\
Our polarized analysis is done using the QCD-PEGASUS package in the
fixed-flavor number scheme with consideration of massless partonic
flavors and $N_{f}=3$ same as unpolarized procedure
\cite{Vogt:2004ns}. Finally our minimization for $\chi_{\mathrm{\rm
global}}^{2}$ is performed with 15 unknown parameters from PPDFs
parametrization forms and we obtain $\frac {\chi^{2}}{NDF}=0.829$
which shows an acceptable fit to the number of 491 experimental
data. Fig.~\ref{pdf} shows the comparison of extracted PPDFs with
other models and the symmetry breaking effect on $\delta\bar{u}$ and
$\delta\bar{d}$ difference, comparing with the results from other
models and experimental data, is presented in Fig.~\ref{mines}.

\section{Summary and conclusions}
\nin
In the present paper we present two NLO QCD analysis of the
unpolarized and polarized data from DIS and SIDIS experiments. While
the analysis we always have $SU(2)$ and $SU(3)$ symmetry breaking
i.e. $\bar{u}\neq \bar{d}\neq\bar s$, but we consider $s=\bar{s}$
since the current available experimental data are not yet enough to
recognize them. The effect of symmetry breaking in determining PDFs
and PPDFs is shown and also we find out that the gluon helicity is
still not well known \cite{DSPIN11}. Having extracted PDFs and
PPDFs, we can determine nucleon unpolarized and polarized structure
functions $F_2$ and $g_1$ too. In general our results are in good
accord with other models determinations and this proves the progress
of the way toward a precise description of the unpolarized and
polarized parton component of the nucleon.
\section*{Acknowledgements}
\nin
 A.N.K thanks CERN TH-PH division for the hospitality where some
 parts of this work was accomplished. The current project was supported financially
by Semnan University and the School of Particles and Accelerators,
Institute for Research in Fundamental Sciences (IPM).






\end{document}